\pdfoutput=1

\documentclass[aps,prb,reprint,showpacs,superscriptaddress,byrevtex]{revtex4-1}

\usepackage{graphicx}
\usepackage{graphics}
\usepackage{amsmath}
\usepackage{amssymb}
\usepackage{amsfonts}
\usepackage{dcolumn}
\usepackage{dsfont}
\usepackage{latexsym}
\usepackage{rotating}
\usepackage{color}
\usepackage{latexsym}
\usepackage{bbm}
\usepackage{subfigure}
\usepackage{float}
\usepackage{epsfig}
\usepackage{epsf}
\usepackage{psfrag}
\usepackage{bm}
\usepackage{amsthm}
\usepackage{eucal}
\usepackage{mathrsfs}
\usepackage{url}
\usepackage{braket}
\usepackage{natbib}

\usepackage{color} 

\usepackage{hyperref}
\hypersetup{
	colorlinks=true,final=true,
	linkcolor=blue,
	citecolor=blue,
	filecolor=blue,
	urlcolor=blue,
}

\begin{document}

\title{Orbital order and electron itinerancy in CoV$_{2}$O$_{4}$ and Mn$ _{0.5} $Co$ _{0.5} $V$_{2}$O$_{4}$ from first principles}
	
\author{Jyoti Krishna}
\affiliation{ Department of Physics, Arizona State University, Tempe, AZ - 85287, USA}
\author{T. Maitra}
\email{tulimfph@iitr.ac.in}	
\affiliation{ Department of Physics, Indian Institute of Technology Roorkee, Roorkee 247667, Uttarakhand, India}

	\begin{abstract}
		
	In view of the recent experimental predictions of a weak structural transition in CoV$_{2}$O$_{4}$ we explore the possible
orbital order states in its low temperature tetragonal phases from first principles density functional theory calculations. We observe that the tetragonal phase with I4$_1/amd$ symmetry is associated with an orbital order involving complex orbitals with a reasonably large orbital moment at Vanadium sites while in the phase with I4$_1/a$ symmetry, the real orbitals with quenched orbital moment constitute the orbital order. Further, to study the competition between orbital order and electron itinerancy we considered Mn$_{0.5}$Co$_{0.5}$V$_{2}$O$_{4}$ as one of the parent compounds, CoV$_{2}$O$_{4}$, lies near itinerant limit while the other, MnV$_{2}$O$_{4}$, lies deep inside the orbitally ordered insulating regime. Orbital order and electron transport have been investigated using first principles density functional theory and Boltzmann transport theory in CoV$_{2}$O$_{4}$, MnV$_{2}$O$_{4}$ and Mn$_{0.5}$Co$_{0.5}$V$_{2}$O$_{4}$. Our results show that as we go from MnV$_{2}$O$_{4}$ to CoV$_{2}$O$_{4}$ there is enhancement in the electron's itinerancy while the nature of orbital order remains unchanged.  
		
	\end{abstract}

	\maketitle
\section{Introduction}
The family of spinel vanadates AV$_{2}$O$_{4}$ (AVO) presents a perfect platform for understanding the interplay among all possible degrees of freedom in a solid such as spin, orbital, lattice and charge \cite{radaelli, Lee2010_5}. In addition, the Mott physics is also at play here, as by mere controlling of the V-V distance (by varying the size of A$ ^{2+}$ ion), the system can be driven to its itinerant state from an orbitally and magnetically ordered insulating state\cite{Kiswandhi2014_5, Kiswandhi_2011}. One can thus observe a fascinating competition between orbital order and electron itinerancy by tuning V-V distance in these materials. An additional geometrical frustration due to pyrochlore lattice of V$^{3+}$ ions, makes spinel vanadates a fertile ground to study competing magnetic interactions (J$_{V-V}$ and J$_{A-V}$) on a geometrically frustrated lattice. Whilst, the partially filled t$_{2g}$ orbitals of V$^{3+}$ ions, makes it always orbitally active, thereby triggering a structural transition, a magnetic transition normally follows at low temperatures due to the partial lifting of geometrical frustration. The multiple transitions (structural and magnetic) in AVO is thus a consequence of the nature of A$^{2+}$ ion being magnetic, orbitally active or both (e.g. MnV$_2$O$_4$, CoV$_2$O$_4$, FeV$_2$O$_4$ etc.)\cite{Garlea2008_5, Kawaguchi2013_5, Kawaguchi2016_5, huang2012_5}. \\

Among the spinel vanadates with magnetic A-site ion, CoV$_{2}$O$_{4}$ is unique in two ways: one, it sits very close to itinerancy limit 2.94 $\AA$ in terms of V-V distance (R$_{V-V}) $\cite{huang2012_5,raman2014_5,reig2016_5} such that on the application of moderate pressure (about $\sim$ 8 GPa) it shows metallic behavior\cite{kisma_5} and two, the existence of any structural transition in this compound is highly debated with most of the earlier experimental reports finding no evidence of cubic to tetragonal structural transition in this system\cite{kisma_5, huang2012_5, Lee_2017} whereas recent reports claim to observe a weak structural transition\cite{Koborinai2016_5,reig2016_5}. Interestingly though, it has got the highest magnetic transition temperatures: a collinear ferrimagnetic transition temperature T$ _{C} $ = 150 K and a non-collinear ferrimagnetic T$_{NC} $ = 90 K\cite{Koborinai2016_5} among the AVOs. Recent strain measurements have reportedly identified a weak first order structural transition ($\Delta a / a \sim 10^{-4}$) at 90 K concurrent with T$ _{NC} $\cite{Koborinai2016_5,reig2016_5}. Further, from the dielectric measurements, the authors proposed an orbital glassy state to be present at low temperatures. From another recent high resolution neutron powder diffraction measurement a very small tetragonal distortion $(1-c/a < 0.06 \%) $ has been captured in CoV$_{2}$O$_{4}$ at 95 K and on further lowering of temperature, a change in the tetragonal symmetry from I4$_1$/amd to I4$_1$/a was observed at 59K via single crystal synchrotron radiation measurement\cite{Ishibashi2017_5}. Further analysis has led the authors to propose an anti-ferro orbitally ordered state below 59K.  \\ 

MnV$_{2}$O$_{4}$, on the other hand, is insulating in character and R$_{V-V}$ is away from the itinerancy limit and orbitally ordered\cite{Nii2012_5,Baek2009_5,Chung2008_5,Garlea2008_5,Zhou2007_5,Adachi2005_5, Suzuki2007_5}. Here, competing superexchange interactions J$_{Mn-V}$ and J$_{V-V}$ lead to two magnetic transitions: a collinear ferrimagnetic transition (T$_{C}$) at 56 K (V spins aligned opposite to that of Mn spins along c-direction) and a non-collinear ferrimagnetic (T$ _{NC} $) transition (V spins cant away from c-axis while Mn spins remain parallel to c-axis) at 53 K. Also, the structural transition (cubic to tetragonal with space group I4$_1$/a) is well established in this system which occurs at 53 K coincident with T$ _{NC} $. In the tetragonal phase, VO$_6$ octahedra are compressed along c-axis because of Jahn-Teller (JT) effect. The tetragonal compression along c-axis lowers the $d_{xy}$ orbital which then gets occupied by one electron. The second electron has therefore the choice of going to either $d_{xz}$, $d_{yz}$ or both paving the way for a possible orbital ordering in the system. The cooperative JT effect forces the orbitals to order in certain fashion relieving the frustration of the lattice.

Many theoretical models are proposed for orbital order (OO) in spinel vanadates that depends on the relative strength of SO coupling, superexchange and JT interaction\cite{Tsunetsugu2003_5,Tchernyshyov2004_5, TM-PRL1}. The possible scenarios for OO in these systems are (1) A-type anti-ferro (real) OO, in which the second electron of V ion, alternately occupies d$_{xz}$/d$_{yz}$ orbital in successive ab-planes along c direction resulting in the quenching of orbital angular momentum L, (2) Ferro (complex) OO, results in the ordering of complex orbitals (d$_{yz}$$\pm$id$_{xz}$) due to enhanced SO coupling leading to unquenched L. X-ray Magnetic Circular Dichroism (XMCD) technique has been very useful for detecting the local orbital moment\cite{Matsuura2015_5,Okabayashi2015_5} which provides the information about real or complex orbital ordering. In case of MnV$_{2}$O$_{4}$, previous DFT calculations indicate real orbital ordering\cite{TM-PRL2,DD-PRB} which was corroborated by XMCD measurements\cite{Matsuura2015_5}.\\

In a recent measurement, Ma et al.\cite{Ma2015_5} have looked into the effect of Co doping in MnV$_{2}$O$_{4}$ and observed a crossover from orbitally ordered low Co-doping regime to increased electron itinerancy in high Co-doped regime. Therefore, it would be interesting to explore using first principles DFT and transport calculations the following two phenomena. Firstly, the nature of orbital ordering in low temperature phases of CoV$_{2}$O$_{4}$ in view of the structural transitions observed recently in experiments\cite{Ishibashi2017_5} and secondly,
the competition between orbital order and electron itinerancy as a function of Co doping as we go from MnV$_{2}$O$_{4}$ to CoV$_{2}$O$_{4}$. We have addressed above two issues in our work presented in this paper. 

\section{Methodology}
The first principles density functional theory and Boltzmann transport theory have been used for the calculations presented in this paper. The structural parameters for MnV$_{2}$O$_{4}$ and CoV$_{2}$O$_{4}$ (in its different structural phases) were obtained from experiments\cite{Nii2012_5,Ishibashi2017_5}. For Mn$ _{0.5}$Co$ _{0.5}$V$_{2}$O$_{4}$,  50$\%$ of Mn ions were substituded by Co in the experimental structure of the MnV$_{2}$O$_{4}$. We have then optimized both parent ( MnV$_{2}$O$_{4}$ and CoV$_{2}$O$_{4}$) and doped structures using 2Doptimize package available in WIEN2k\cite{Blaha_5}. This package effectively performs 2D optimization in volume, $c/a$ and atomic positions. The optimization (volume, $c/a$ and oxygen positions) were carried out within Perdew-Burke-Ernzerhof Generalized Gradient Approximation (PBE-GGA)\cite{Perdew_5} exchange-correlation functional within the full potential linearized augmented plane wave (FP-LAPW) method as implemented in WIEN2k. In this, we have considered nine different structures having their volumes varied by 0.5\%, 1.0\%, 1.5\% and 2.0\% with respecto to experimental structure, and for each volume, nine different $c/a$ structures were taken. We obtained the minimum energy structure from a total 81 different structures where oxygen positions were allowed to relax. Using this optimized structure, we performed further self-consistent field calculations considering 78 $\vec{k}$ points in the irreducible Brillouin zone and plane wave cut off (R$_{mt}$K$_{max}$) to be 8.0. The muffin tin radii were taken as 2.0, 1.95 and 1.65 a.u. for Mn/Co, V and O respectively. Due to the presence of 3d electrons, Coulomb correlation (U) becomes indispensable, which was included in our calculations within GGA+U\cite{Anisimov1993_5}. This takes into account the on-site Coulomb interaction and removes the self Coulomb and self exchange-correlation energy. The spin-orbit coupling (SO) is included by the second variational method with scalar relativistic wavefunctions\cite{koelling_5}. To analyse the orbital order and electron transport we used maximally-localized Wannier functions (MLWF) to fit the DFT bands calculated by WIEN2k. WANNIER90 \cite{WANNIER_5} and WIEN2WANNIER\cite{Kunes_5} codes were used for this purpose. We calculated the transport properties using BoltzWann code that utilizes semi-classical Boltzmann transport theory\cite{boltz_5}.

\section{Results and Discussions}	
\begin{table}[]
	\centering
	\caption{\label{TABLE I.} The V-O bond lengths in the ab-plane and along c, A-O bond length for x=0.0 (MnV$_{2}$O$_{4}$), x=0.5 (Mn$_{x}$Co$_{1-x}$V$_{2}$O$_{4}$) and x=1.0 (CoV$_{2}$O$_{4}$) after structural optimization .}
	\renewcommand{\arraystretch}{1.5}
	\begin{tabular}{||c|c|c|c||}
		\hline
		\textbf{x}& \textbf{V-O (c)}& \textbf{V-O (ab)}& \textbf{A-O} \\ 
		\hline\hline
		\textbf{0.0}&2.0293&(2.0434,2.0137) & 2.0346 \\ 
		\hline
		\textbf{0.5}&2.0191&(2.0331,2.0036) & 2.0244\\
		\hline
		\textbf{1.0}&&& \\
		$Fd3m$ (PHASE 1)&2.0169&(2.0169,2.0169)& \\
		$I41/amd$ (PHASE 2)&2.0148&(2.0176,2.0176)& \\
		$I41/a$ (PHASE 3)&2.0150&(2.0059,2.0295)& 1.9696 \\
		\hline
	\end{tabular}
\end{table} 
In view of recent experimental report\cite{Ishibashi2017_5} on the observation of weak structural transitions in CoV$_{2}$O$_{4}$ from cubic (Fd$\bar{3}$m) to tetragonal (I4$_1$/amd and I4$_1$/a) phases, we investigated the electronic structure and nature of orbital ordering in the experimentally observed tetragonal phases in this system from first principles calculations. Further, we also looked into the effect of Co doping on the competition between orbital ordering and electron itinerancy as we go from  MnV$_{2}$O$_{4}$ to CoV$_{2}$O$_{4}$.
The results presented in this section have therefore been divided into two parts. In part A, we present our calculations and results on the electronic structure and orbital order analysis of the cubic (Fd$\bar{3}$m) and tetragonal (I4$_1$/amd and I4$_1$/a) phases of CoV$_{2}$O$_{4}$. In part B, we present our results on the effect of Co doping (x) on MnV$_{2}$O$_{4}$ where we have discussed electronic structure,
orbital ordering and electron transport calculations for the parent and doped compounds. Electronic structure calculations are performed within GGA+U and GGA+U+SO approximations with U$_{eff}$ =U-J= 4 eV for Mn/Co and 3 eV for V ions where U is the Coulomb correlation and J is Hund's coupling. 
\begin{figure}[ht!]
	\centering
	\includegraphics[scale=0.3]{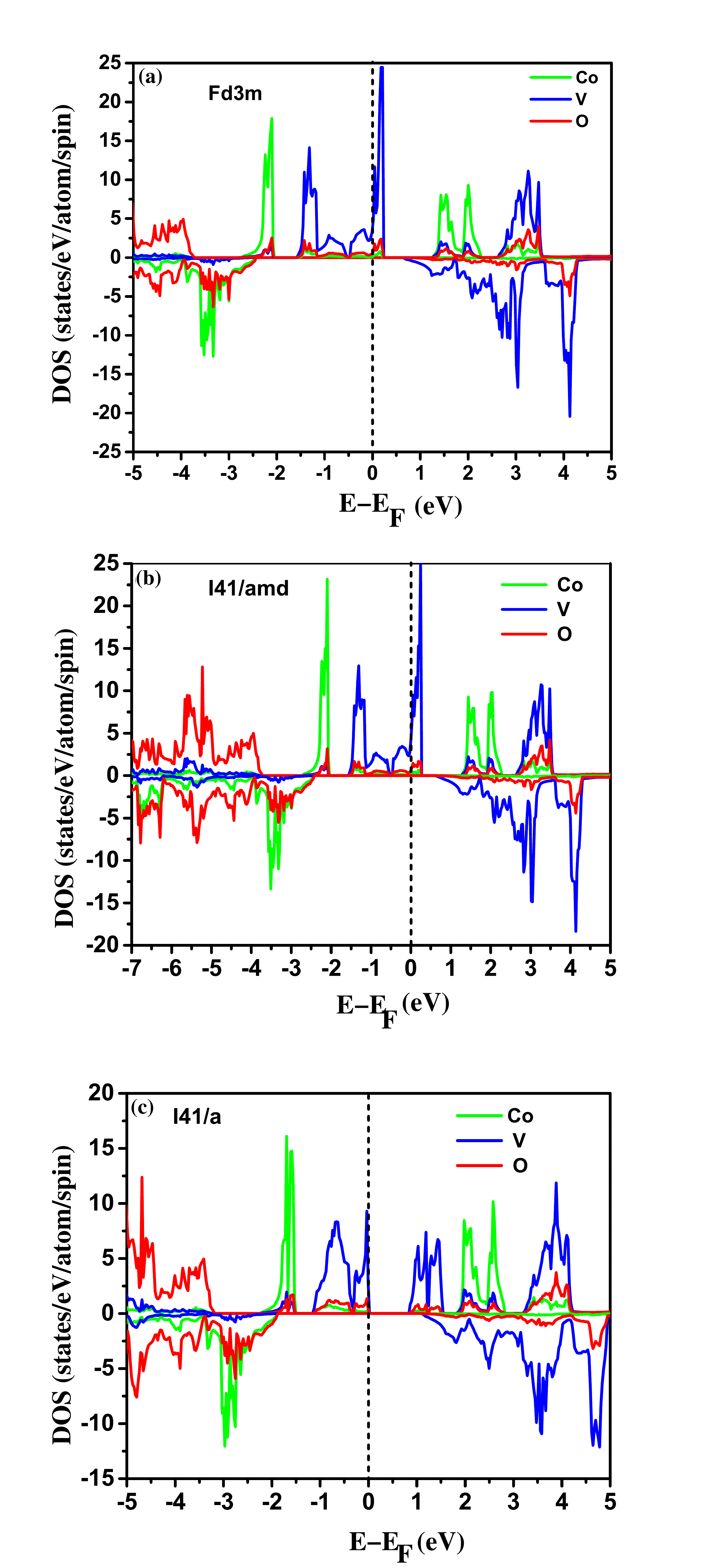}
	\caption{\label{fig: 1} The total density of states (DOS) for Co, V, and O within GGA+U for (a) cubic Fd$\bar{3}$m, (b) tetragonal $I4_1/amd$ and (c) $I4_1/a$ phases in majority and minority spin channel. }
\end{figure}   

\begin{figure*}[!ht]
	\centering
	\includegraphics[scale=0.3]{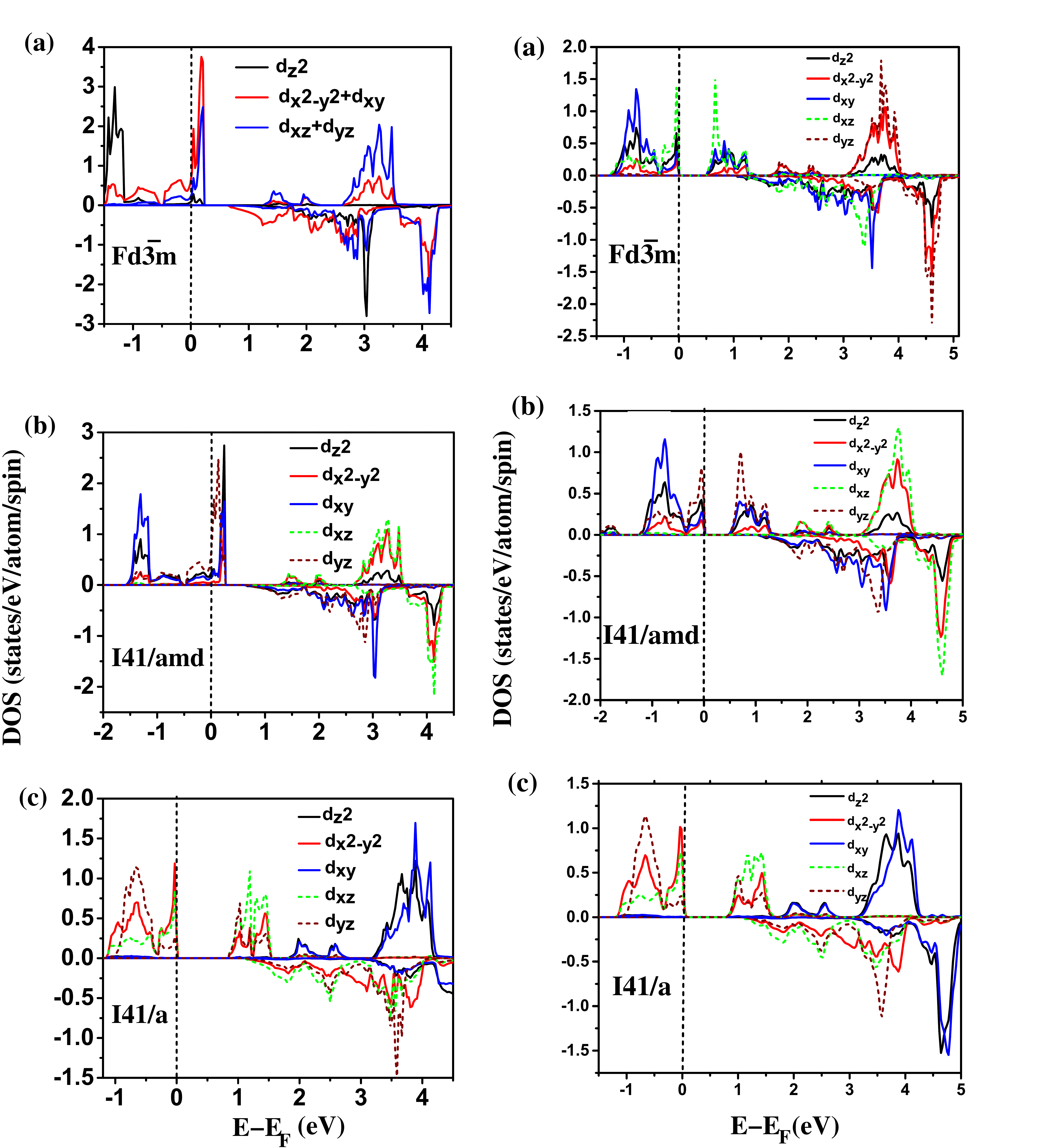}
	\caption{\label{fig: 2} The Vanadium partial density of states (DOS) for $d$-electrons within GGA+U (left) and for GGA+U+SO (right) for (a) cubic Fd$\bar{3}$m, (b) tetragonal $I41/amd$ and (c) $I41/a$ phases for majority and minority spin channels. Note that the set [d$_{x^2-y^2}$, d$_{xz}$ and d$_{yz}$] form t$_{2g}$ rather than conventionally used [d$_{xy}$, d$_{xz}$ and d$_{yz}$] as the crystallographic $a$ and $b$-axes are rotated by 45$^{\circ}$ with respect to planar V-O bonds of VO$_6$ octahedra.}
\end{figure*}   
\begin{figure*}[!ht]
	\centering
	\includegraphics[width=14cm]{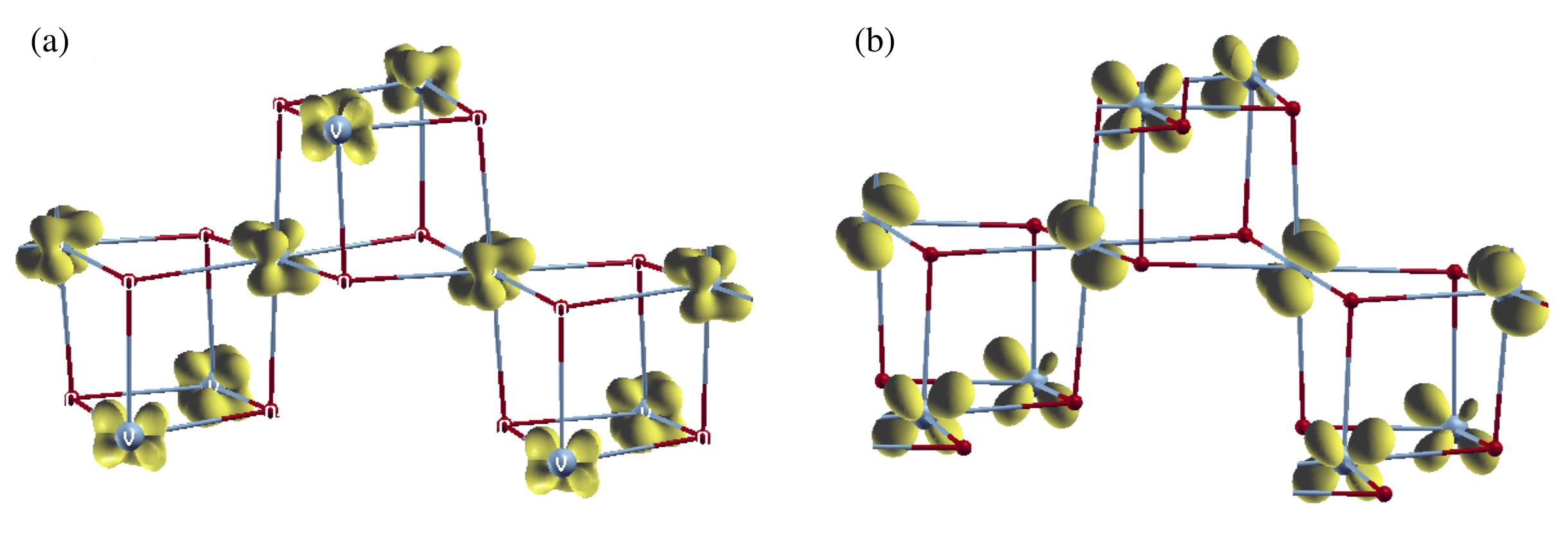}
	\caption{\label{fig: 3} 3D hole density in real space depicting the nature of unoccupied orbitals of V$^{3+}$ ion for (a) phase 2 and (b) phase 3. The rotation of the orbitals (in ab plane and c-direction) in the V chain is because of the trigonal distortion. }
\end{figure*} 

\subsection{CoV$_{2}$O$_{4}$ and its different phases}

The GGA+U calculations were carried out on the optimized structures (see Table I) for all the three structural phases (cubic (phase 1), tetragonal I4$_1$/amd (phase 2) and tetragonal I4$_1$/a (phase 3)) by considering a collinear ferrimagnetic spin configuration where the spins of V are antiparallel with respect to Co spins along c-axis. The total energy calculations reveal that the tetragonal phase with I4$_1$/a symmetry has the lowest energy which is consistent with the experiment. In Figure \ref{fig: 1} we present the total density of states (DOS) of all the three phases of CoV$_2$O$_4$ for both the spin channels calculated within GGA+U approximation. While the cubic and tetragonal (I4$_1$/amd) phase display the metallic character even with large $U$ (in contrary to the experiments), an insulating behavior is observed for tetragonal phase 3 (I4$_1$/a) (consistent with experiments). 
From DOS one can clearly observe that the tetrahedral crystal field splits the d-states of Co$^{2+}$ ions into the lower energy e$ _{g}$ and higher energy t$ _{2g}$ states. Four out of the seven valence electrons of Co$^{2+}$ thus completely fill the e$ _{g}$ states (both majority and minority) and the rest three electrons make t$ _{2g} $ states half-filled (only majority states are filled, minority states remain empty). Because of this, the filled Co d-states lie much below the Fermi level (FL). On the other hand,  V$ ^{3+} $ ions are octahedrally coordinated by O ions thus the d-states here split into higher energy e$ _{g}$ and lower t$ _{2g}$ states. The two d-electrons of V$ ^{3+}$ ions thus partially fill the t$ _{2g}$ states that lie near Fermi level (FL). Tetragonal distortions present in phase 2 does not split the t$ _{2g}$ states completely to open a gap while in phase 3 a finite gap is seen in DOS due to the splitting of t$ _{2g}$ states (see Figure \ref{fig: 2} below). Note that in addition to octahedral crystal field, the trigonal distortion is also present in case of V ions (as O-V-O angles deviate from 90$^{\circ}$). Because of varying strengths of tetragonal and trigonal distortions present in three phases the nature and magnitude of splitting of t$ _{2g}$ states are different. 
\begin{figure*}[!ht]
	\centering
	\includegraphics[scale=0.35]{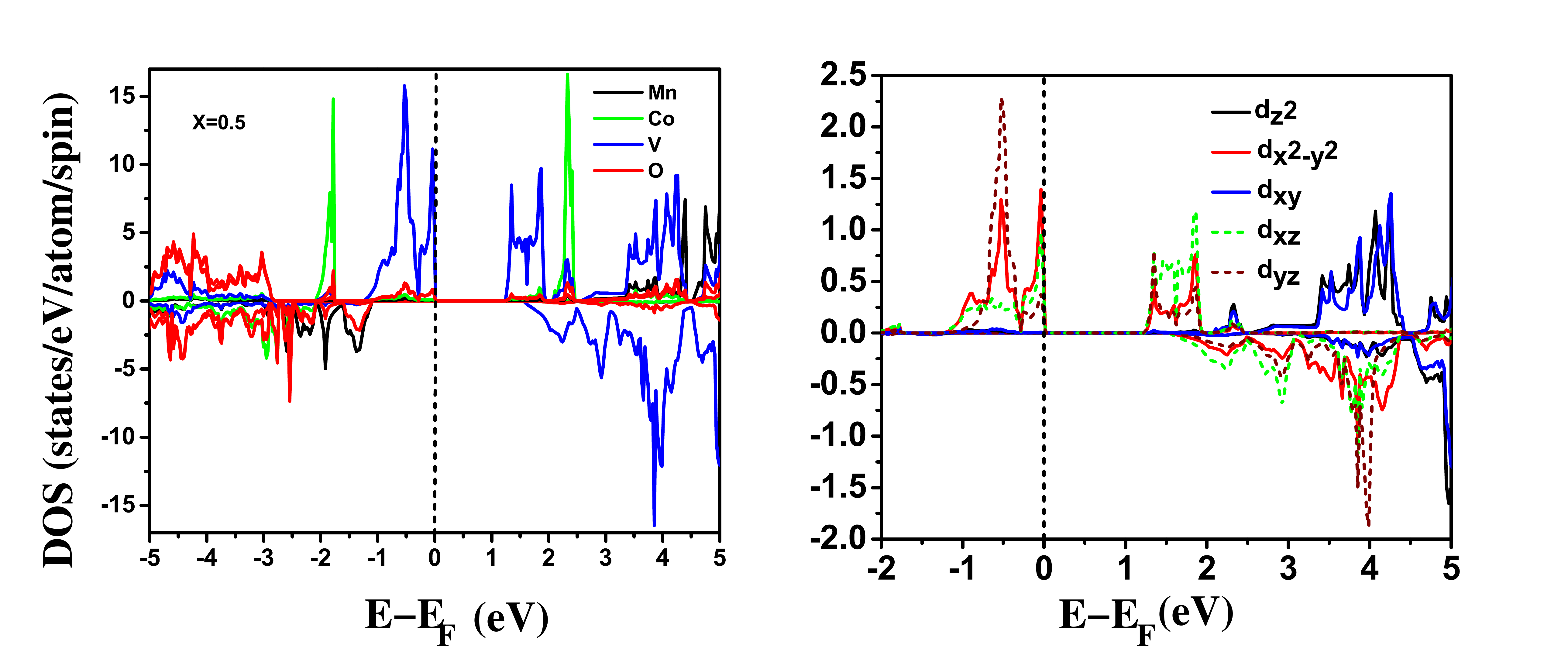}
	\caption{\label{fig: 4} \textit{Left}: The spin-polarized total DOS for 50\% (x=0.5) Co doping at MnV$_{2}$O$_{4}$ site calculated within GGA+U approximation. \textit{Right} The vanadium partial DOS for \textit{d} states.}
\end{figure*} 

We present in Figure \ref{fig: 2}, the partial V DOS of three different phases calculated within GGA+U (left panel) and GGA+U+SO (right panel). One can see from Figure \ref{fig: 2} (left) that in phase 1 (the cubic phase) where only trigonal distortion is present, the t$_{2g}$ states split into a lower singlet a$_{1g}$ and the higher doublet e$_{g}^{\prime}$ states which are linear combinations of d$_{xy}$, d$_{xz}$ and d$_{yz}$ orbitals. In phase 2, the presence of both tetragonal and the trigonal distortion create much complicated splitting of t$_{2g}$ states. In phase 3, the extent of trigonal distortion is much lower than the tetragonal compression thus t$_{2g}$ split into pure d$_{xy}$, d$_{xz}$ and d$_{yz}$ states. 
In phase 1 and 2, the highest occupied t$ _{2g}$ level remains doubly degenerate and filled by one electron and hence the system shows metallic behavior. In phase 3, however, we see a further splitting of the t$ _{2g}$ levels giving rise to a band gap in the DOS as seen in Figure \ref{fig: 2}(c)(left). This splitting is because of the Jahn-Teller (JT) distortion associated with I4$_1$/a symmetry where the V-O bonds are different in all three directions. This is also associated with a long range orbital ordering (OO) as discussed below. Comparing the DOS of GGA+U and GGA+U+SO we observe that for phase 1 and 2 they are drastically different whereas for phase 3 they are very simillar. In phase 1 and 2 we obtain an insulating electronic structure within GGA+U+SO, unlike the metallic solution of GGA+U, which is consistent with experiment. The SO effect on DOS is seen to be quite significant in phase 1 and 2, whereas in phase 3 it is negligible. This is also reflected in calculated orbital moment values at V site which are -0.66 $\mu_B$ and -0.06 $\mu_B$ for phase 2 and 3 respectively. 

In order to understand the nature of OO, which is difficult to ascertain from the DOS, we have calculated 3D electron and hole densities in real space at V sites and have also done an analysis using tight-binding (Wannier orbitals) basis for phase 2 and 3. In Figure \ref{fig: 3} we show the calculated 3D hole densities for phase 2 and phase 3 at V sites within GGA+U+SO approximation. Comparing these two one can clearly see that that unoccupied V orbital structure and hence the occupied ones are very different in phase 2 and 3. As also evident from the large orbital moment (-0.66$\mu_B$) value in case of phase 2, we propose the presence of a complex orbital order of the type $d_{xz}\pm id_{yz}$ in this case as seen in ZnV$_2$O$_4$\cite{TM-PRL1}. While in phase 3, a further analysis using Wannier orbitals basis alongwith a small value of calculated orbital moment (-0.06$\mu_B$) support A-type orbital order where real orbitals d$_{yz}$ and d$_{xz}$ are occupied at V sites in successive ab planes along c-direction. Therefore, we conclude that the transition seen
in experiment\cite{Ishibashi2017_5} at 59K where the crystalline symmetry changes from I4$_1/amd$ to I4$_1/a$ is also associated with complex to real orbital order transition. Future XMCD measurements are expected to shed further light on this.

\begin{figure}[h]
	\centering
	\includegraphics[scale=0.4]{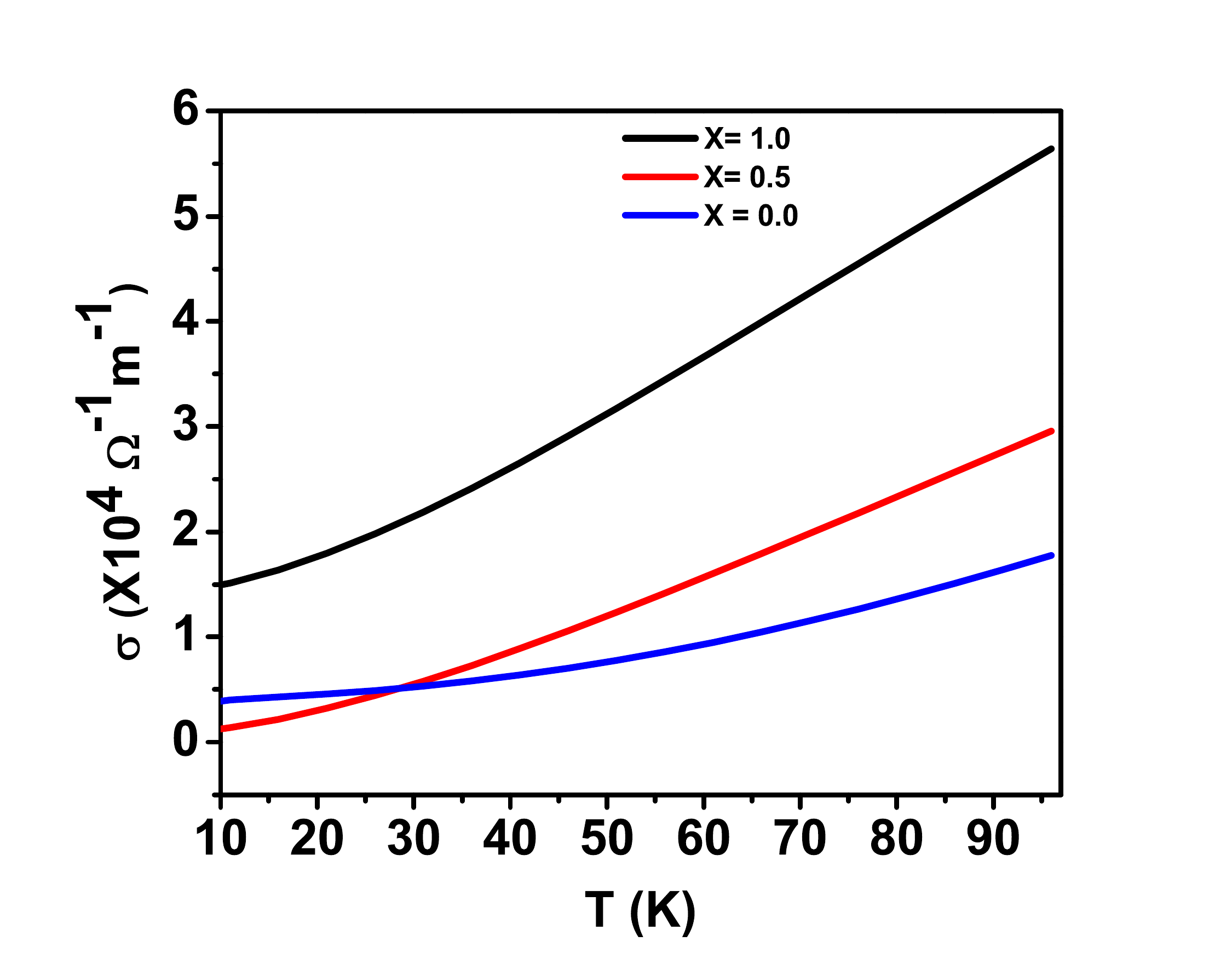}
	\caption{\label{fig: 5} The temperture dependence of electrical conductivity ($\sigma$) for different doping concentrations (x) in Mn$_{1-x}$Co$_x$V$_2$O$_4$. }
\end{figure} 
 
\subsection{50$\%$ Co doped MnV$_{2}$O$_{4}$ } 
As discussed in the introduction, MnV$_{2}$O$_{4}$ is known to be orbitally ordered while CoV$_{2}$O$_{4}$ is close to the itinerant regime. In view of the recent experimental measurements by Ma et al.\cite{Ma2015_5} and to study the competition between electron itinerancy and orbital order in Co doped MnV$_{2}$O$_{4}$ from theoretical perspective, we considered 50$\%$ of Mn by Co in MnV$_{2}$O$_{4}$(i.e. Mn$_{1-x}$Co$_x$V$_2$O$_4$ with x=0.5). In a previous study we have reported the effect of Co doping on the structural parameters such as lattice constants, inter-vanadium distance (R$_{V-V}$) etc.\cite{jkconf_5}.  With increase in Co doping at Mn sites the lattice constants $a$ and $c$ are seen to decrease. This is because of the smaller size of Co ion compared to that of Mn. This induced chemical pressure due to Co doping also affects R$_{V-V}$. We also observe that with Co doping, the Mn-O bond lengths along c-axis decrease due to the hydrostatic pressure acting at Mn site which shrinks Mn-O tetrahedron (see Table I). This contraction enhances local trigonal distortion in VO$_{6}$ octahedron.
\begin{figure*}
	\centering
	\includegraphics[width=18cm]{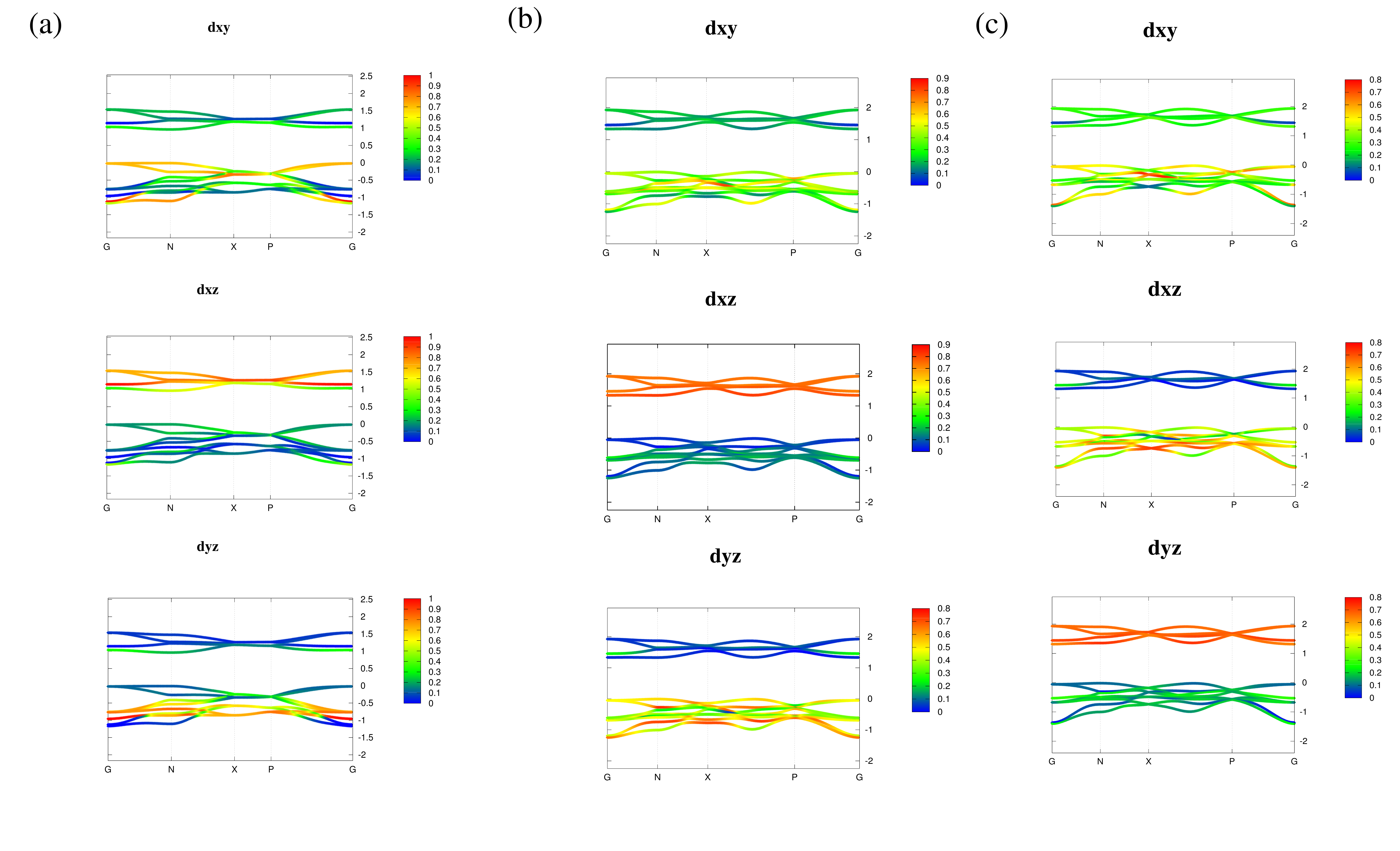}
	\caption{\label{fig: 6} Projection of V t$_{2g}$ bands on to d$_{xy}$, d$_{xz}$ and d$_{yz}$ Wannier orbitals for (a) CoV$_{2}$O$_{4}$ in phase 3 (b) MnV$_{2}$O$_{4}$ and (c) x=0.5. The color bar on right side depicts the orbital character with the red (blue) color showing highest (lowest) character of a particular orbital.}
\end{figure*}

In Figure \ref{fig: 4} we present the total and V partial DOS of Mn$_{0.5}$Co$_{0.5}$V$_{2}$O$_{4}$ within GGA+U approximation. Comparing the DOS for this doped compound with the same for parent compounds CoV$_{2}$O$_{4}$ (phase 3, see Figure \ref{fig: 2}) and MnV$_{2}$O$_{4}$ (not shown here), we observe that the bandgap decreases. To clearly establish that the electron itinerancy indeed increases with Co doping we have calculated the electrical conductivity as a function of temperature using Boltzmann transport theory. For this we have used Wannier fits to DFT bands of V t$_{2g}$ states around FL obtained from Wien2k within GGA+U. The variation of electrical conductivity ($\sigma$) with temperature for MnV$_{2}$O$_{4}$ (x=0.0), CoV$_{2}$O$_{4}$ (x=1.0) and 50$\%$ Co doped (x=0.5) MnV$_{2}$O$_{4}$ are shown in Figure \ref{fig: 5}. We can clearly see that $\sigma$ rises as Co concentration increasing indicating an enhancement of electron itinerancy in the system. Also, we see that with increasing temperature $\sigma$ rises in all three cases showing the insulating character of the systems. 

To ascertain how much orbital order is affected by Co doping, we have performed Wannier analysis of the V t$_{2g}$ bands around the FL of CoV$_{2}$O$_{4}$, MnV$_{2}$O$_{4}$ and x = 0.5 Co doped case. The comparison is presented in Figure \ref{fig: 6}. The projections of individual Wannier d orbitals on to V d-bands of the respective compounds are shown. The color code depicts the strength of the orbital character with red representing the highest. One can clearly see that in all three cases d$_{xy}$ orbital is predominantly occupied at a V site whereas there is a difference between occupancies of d$_{xz}$ and d$_{yz}$ orbitals. Further analysis by plotting these Wannier orbitals in real space reveals that A-type OO similar to phase 3 of CoV$_2$O$_4$ prevails in all three compounds (i.e. one electron of V$ ^{3+}$  always occupies at d$_{xy}$ orbital with another electron alternating between d$_{xz}$ and d$_{yz}$). 
If we consider the difference between the orbital occupancies of d$_{xz}$ and d$_{yz}$ orbitals as the strength of the orbital order, then one can clearly see from Figure \ref{fig: 6} that in MnV$_{2}$O$_{4}$ and x = 0.5 Co doped casethis difference is somewhat larger that the same in CoV$_{2}$O$_{4}$. Thus one can conclude from the above discussion that as we go from MnV$_{2}$O$_{4}$ to CoV$_{2}$O$_{4}$, the electron itinerancy increases, the nature of orbital order remains same while the strength of orbital order weakens.  
Finally we would like to mention that we have also performed calculations including SO interaction for the doped compound to see whether complex orbitals are present in this case. The orbital moment for V ion is found to be -0.068 $\mu_{B}$ which is very small. Thus here also only real orbitals are involved in OO.

\section{Conclusions}
We have performed the first principles calculation in the three different structural phases of CoV$_{2}$O$_{4}$: cubic, tetragonal with I4$_1$/amd symmetry and tetragonal with I4$_1$/a symmetry. Total energy calculations reveal that tetragonal with I4$_1$/a symmetry is indeed lower in energy with respect to the cubic phase. Our GGA+U calculations show that while cubic and tetragonal phase with I4$_1$/amd symmetry remain metallic in character, the tetragonal phase with I4$_1$/a symmetry becomes insulating. We observe further from our GGA+U+SO calculations that tetragonal phase with I4$_1$/amd symmetry has orbital order involving complex orbitals with large orbital moment whereas tetragonal phase with I4$_1$/a symmetry has A-type orbital order comprising of real orbitals with negligible orbital moment. Therefore, we conclude that the structural transition seen in experiment at 59K where the symmetry of tetragonal phase changes from I4$_1$/amd to I4$_1$/a, is associated with complex to real orbital order transition. We also performed calculations for 50\% Co doped MnV$_2$O$_4$ compound (i.e. Mn$_{0.5}$Co$_{0.5}$V$_{2}$O$_{4}$) to study the competition between electron itinerancy and orbital order as a function of doping. We observed that doping Co at Mn sites indeed reduces the strength of orbital order even though the orbital order remains the same. At the same time, it enhances the electron itinerancy which we could clearly see from calculated electrical conductivity which increases with the Co content.
  
\section{Acknowledgement}
This work is supported by DST-DAAD project (grant no INT/FRG/DAAD/P-16/2018). JK acknowledges MHRD(INDIA) for research fellowship.

  \end{document}